\documentclass[%
 reprint,
 amsmath,amssymb,
 aps,
]{revtex4-1}
\usepackage{graphicx}
\usepackage{dcolumn}
\usepackage{bm}
\usepackage{graphicx}
\usepackage{dcolumn}
\usepackage{bm}
\usepackage{bigints}
\usepackage{color}
\usepackage{amsmath,amssymb,graphicx}
\usepackage{float}
\usepackage{braket}
\usepackage{tabularx}
\usepackage{tabulary}
\usepackage{tabu}
\usepackage{booktabs}
\usepackage{textcomp}

\newcolumntype{C}{>{\centering\arraybackslash}X}

\begin{document}
\preprint{APS/123-QED}
\title{Advances in Disordered Transverse Anderson Localizing Optical Fibers}

\author{Arash Mafi$^{1,*}$}%
\author{Matthew Tuggle$^{2}$}%
\author{Cody Bassett$^{1}$}%
\author{Esmaeil Mobini$^{1}$}%
\author{John Ballato$^{2}$}%

\email{mafi@unm.edu}
\affiliation{$^1$Center for High Technology Materials (CHTM)  and the Department 
of Physics \& Astronomy, University of New Mexico, Albuquerque, NM 87131, USA \\
             $^2$Center for Optical Materials Science and Engineering Technologies (COMSET) and the Department 
of Materials Science and Engineering, Clemson University, Clemson, SC 29625, USA} 

\begin{abstract}
Disordered transverse Anderson localizing optical fibers have shown great promise in various applications from image transport 
to random lasing. Their success is due to their novel waveguiding behavior, which is enabled by the transverse Anderson localization 
of light. The strong transverse scattering from the transversely disordered refractive index structure results in transversely confined 
modes that can freely propagate in the longitudinal direction. Therefore, these fibers behave like large-core highly multimode
optical fibers, with the peculiar property that most modes are highly localized. This property makes them ideal for such applications as image transport
and spatial beam multiplexing. 
In this review paper, we will explore some of the recent advances in these fibers, especially those related to the material structure
and fabrication methods. 
\end{abstract}
\maketitle
\section{Introduction}
Transverse Anderson localizing optical fibers (TALOF)~\cite{Mafi-Salman-OL-2012} are a novel class of optical fibers that guide light, not in a conventional core-cladding setting, but using the Transverse Anderson localization (TAL), where any location across the transverse profile of the fiber can be used to guide light. Anderson localization (AL) is the absence of diffusive wave transport in highly disordered scattering media~\cite{Anderson1,Lagendijk-Physics-Today-2009,sheng2006introduction} and is broadly applicable to quantum and classical waves~\cite{Soukoulis-1999,matter-waves-2008,Lahini-Quantum-Correlation-2010,Abouraddy-entangled-2012,ultrasound-1990,acoustic-PRL-1990,elastics-Nat-Phys-2009,John-EM-abs-mobility-edge-1984,Chabanov-microwave-2000,El-Dardiry-microwave-2012,Anderson2,John-Physics-Today-1991,SegevNaturePhotonicsReview,Mafi-AOP-2015}. Transverse Anderson localization (TAL), which is essentially AL in a transversely disordered and longitudinally invariant optical system, was first proposed in a pair of visionary theoretical papers by Abdullaev \textit {et al}.~\cite{transverse-Abdullaev} in 1980 and De~Raedt \textit {et al}.~\cite{transverse-DeRaedt} in 1989. In TAL structures, the dielectric constant is uniform along the direction of the propagation of light, similar to a conventional optical waveguide, and the disorder in the dielectric constant resides in the (one or two) transverse dimension(s). An optical field that is launched in a TAL structure undergoes a brief expansion in the transverse dimension(s) as it propagates freely in the longitudinal direction; however, the strong transverse scattering eventually halts the expansion, and the transversely localized field propagates along the structures just like a conventional optical waveguide~\cite{Mafi-AOP-2015}.    

TAL is ubiquitous in unbounded transversely disordered optical structures; however, to observe TAL for finite transverse size (similar to conventional optical fibers), the scattering due to the disorder must be sufficiently strong such that the transverse physical dimensions of the waveguide are larger than the transverse localization length. Of course, it is assumed that the waveguide is uniform in the longitudinal direction. Over the years, several experiments have successfully observed TAL, the earliest of which are by Schwartz \textit {et al}.~\cite{Schwartz2007} in a photorefractive crystal, and by Lahini \textit {et al}.~\cite{Lahini-1D-AL-2008} in a one-dimensional (1D) lattice of coupled optical waveguides patterned on an AlGaAs substrate. In this paper, we will briefly review the development of TALOFs and outline the best practices regarding the choice of materials and microstructures that can facilitate specific applications. Discussions on the applications of TALOFs, especially for imaging and random lasing can be found in Refs.~\cite{Mafi-AOP-2015,mafi-JLT-ArXiv-2019,Mafi-Salman-Nature-2014,Mafi-Behnam-Random-Laser-2017} and will also be discussed briefly here.  

\begin{figure}[htp]
  \centering
  \includegraphics[width=0.8\columnwidth]{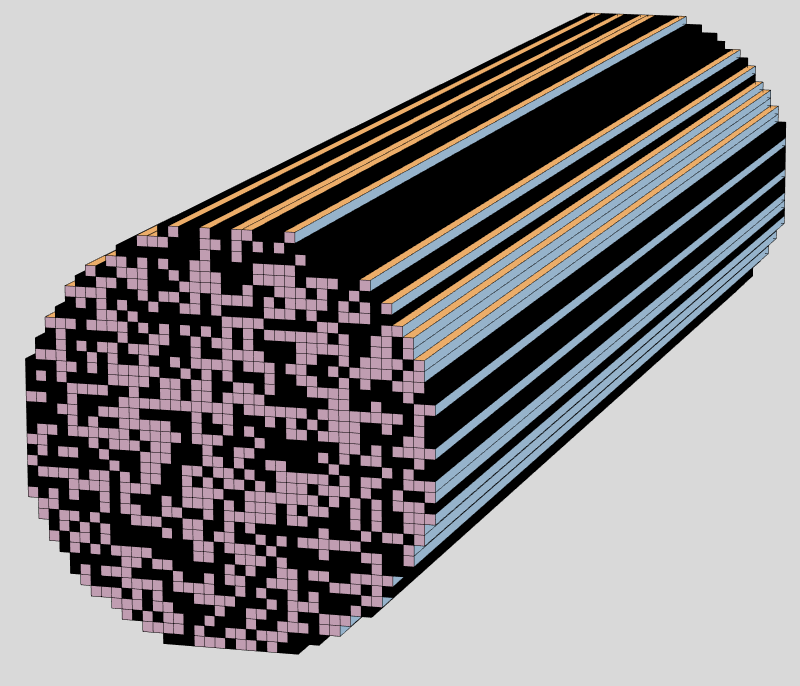}
\caption {A conceptual sketch of the longitudinally invariant and transversely random dielectric
waveguide is shown. This structure which was proposed by De~Raedt \textit {et al}.~\cite{transverse-DeRaedt}
is the basis of TALOFs discussed in this paper. In the transverse plane, the refractive index is 
pixelated into tiny squares, and the refractive index of each pixel is randomly selected to be 
$n_1$ or $n_2$ with equal probabilities.
}
\label{fig:rae-array}
\end{figure}

The TALOF designs are mainly based on the structure proposed by De~Raedt \textit {et al}.~\cite{transverse-DeRaedt}, which is sketched in Fig.~\ref{fig:rae-array}.
The optical medium in Fig.~\ref{fig:rae-array} consists of an optical fiber-like structure, whose refractive index profile is invariant in the longitudinal direction. In the transverse plane, the refractive index is pixelated into tiny squares, where the refractive index of each pixel is randomly selected to be $n_1$ or $n_2$ with equal probabilities. The edge length of each square is on the order of the wavelength of the light. De~Raedt \textit {et al}. showed that an optical field that is launched in the longitudinal direction remains localized in the transverse plane due to the transverse scattering, as it propagates freely in the longitudinal direction. The localization radius can be generally reduced by increasing the transverse scattering strengths, which is an increasing function of the refractive index contrast $\Delta n=|n_2-n_1|$~\cite{transverse-DeRaedt,Mafi-Salman-OPEX-2012,Mafi-AOP-2015}. 
\section{Transverse Anderson localizing optical fibers: existing designs}
\begin{figure}[h]
  \centering
  \includegraphics[width=\columnwidth]{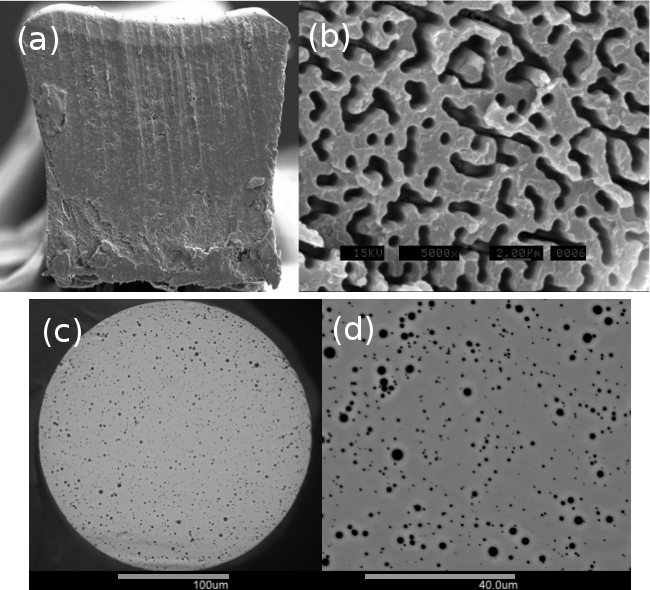}
\caption {(a) Cross section of the polymer TALOF from Ref.~\cite{Mafi-Salman-OL-2012} with a 
nearly square profile and an approximate side-width of 250\,\textmu m,
(b) A zoomed-in SEM image of a 24\textmu m-wide region on the tip of the fiber, exposed to a solvent 
to differentiate between PMMA and PS polymer components, (c) SEM image of the glass TALOF reported in
 Ref.~\cite{Mafi-Salman-OMEX-2012}, (d) A zoomed-in SEM image of the glass TALOF. Reprinted/Reproduced
with permission from Optics Letters, 2012~\cite{Mafi-Salman-OL-2012} and  
Optical Material Express, 2012~\cite{Mafi-Salman-OMEX-2012}, and the Optical Society.}
\label{fig:polymer-glass}
\end{figure}

The first demonstration of TAL in an optical fiber was reported in 2012 by Karbasi \textit {et al}.~\cite{Mafi-Salman-OL-2012}.
The TALOF is shown in Fig.~\ref{fig:polymer-glass}(a)--it was fabricated by the stack-and-draw method
from a low-index component, polymethyl methacrylate (PMMA) with a refractive index of 1.49, and a high-index component,
polystyrene (PS) with a refractive index of 1.59 ($\Delta n=0.1$). 40,000~pieces of PMMA and 40,000 pieces of PS fibers were 
randomly mixed~\cite{Mafi-Salman-JOVE-2013}, fused together, and redrawn to a fiber with a nearly square profile and 
approximate side-width of 250\,\textmu m. Figure~\ref{fig:polymer-glass}(b) shows the zoomed-in scanning electron microscope~(SEM) 
image of an approximately 24\,\textmu m-wide region on the tip of the fiber after exposing the tip to an ethanol solvent to dissolve 
the PMMA. The typical random feature size in the structure shown in Fig.~\ref{fig:polymer-glass}(b) is around 0.9\,\textmu m.
The TALOF was subsequently used for detailed studies of TAL~\cite{Mafi-Salman-OPEX-2012}, 
spatial beam multiplexing~\cite{Mafi-Salman-Multiple-Beam-2013}, and image transport~\cite{Mafi-Salman-Nature-2014}.

The first successful fabrication of a silica TALOF was reported by Karbasi \textit {et al}.~\cite{Mafi-Salman-OMEX-2012} 
in~2012. The glass-air disordered fiber was drawn at Clemson University from ``satin quartz'' (Heraeus Quartz) which
is a porous artisan glass. The starting rod had dimensions of 8\,mm in diameter and 850\,mm in length and was drawn
at a temperature of 1890\textdegree{}C to a fiber with the diameter of 250\,\textmu m. The cross-sectional SEM image of this fiber is shown in 
Fig.~\ref{fig:polymer-glass}(c), and a zoomed-in SEM image is shown in Fig.~\ref{fig:polymer-glass}(d). 
The light-gray background matrix is glass, and the random black dots represent the airholes, where the diameters of the 
airholes vary between 0.2\,\textmu m and 5.5\,\textmu m. Although the air-glass design with $n_1=1.0$ and $n_2=1.46$ ($\Delta n=0.46$)
led to a sharp index contrast and consequently a strong transverse scattering, the fill-fraction of the airholes in this fiber
ranged from nearly 2\% in the center of the fiber to approximately 7\% near the edges; therefore, TAL was only 
observed near the periphery of the fiber. This fiber was subsequently used for the first 
observation of random lasing in a TALOF~\cite{Mafi-Behnam-Random-Laser-2017}.

In 2014, there was another successful attempt in observing TAL in an air-glass optical fiber by Chen and Li~\cite{chen2014observing} at
Corning Incorporated. They fabricated random air-line fibers using the outside vapor deposition process by first creating a pure silica 
soot blank by soot deposition in laydown process. After laydown, the silica soot blank was chlorine dried first at 
1125\textdegree{}C for one hour in a consolidation furnace. Then the blank was consolidated at 1490\textdegree{}C in the presence of 
100\% ${\rm N_2}$. During the sintering process, ${\rm N_2}$ was trapped in the blank to form glass with randomly distributed air bubbles.
The random air bubbles were subsequently drawn into random airlines when the preform was drawn to three fiber samples 
with approximately 150, 250 and 350\,\textmu m diameters. The airline diameters ranged from 0.18\,\textmu m to 0.39\,\textmu m
depending on the fiber diameter, and the air-fill fraction was measured to be around 1\%, which is a significantly lower air 
fill-fraction than those reported in Ref.~\cite{Mafi-Salman-OMEX-2012}. However, the authors reported the successful observation 
of TAL. This can be attributed to the far-subwavelength size of the transverse scattering centers and the higher scatterer 
density (air-line density) compared to Ref.~\cite{Mafi-Salman-OMEX-2012}. 

Recently in 2018, Zhao \textit {et al}.~\cite{ZHAO:17,zhao2018image} at CREOL, University of Central Florida, reported a
glass-air TALOF that was fabricated using the stack-and-draw technique. They randomly mixed thousands of silica 
capillaries with different diameters and airhole diameters and assembled them in a jacket tube to make the preforms. 
Subsequently, fibers were drawn to desired diameters. For example, they reported drawing a fiber with the inner diameter of the 
randomly disordered region at 278\,\textmu m, and the outer diameter of 414\,\textmu m. The air-filling fraction was about 28.5\%
and the air-hole areas ranged from 0.64\,\textmu m$^2$ to over 100\,\textmu m$^2$, with 2.5\,\textmu m$^2$ being the peak of the 
statistical distribution. The fiber was used for the first successful report of high-quality optical image transfer through a 
90 cm-long TALOF. In a more recent attempt, Zhao \textit {et al}.~\cite{zhao2018deep} applied deep learning techniques
to improve the quality of the image transport in these fibers.

Another successful demonstration of TALOFs was reported in 2019 by Tuan \textit {et al}.~\cite{tong2018characterization} from the
Toyota Technological Institute in Japan. This effort is notable for two reasons: first, it is an all-solid structure, so two
species of glass are used to create the random structure; and second, the fabrication is based on tellurite optical glasses 
for applications in near-infrared image transport. To avoid undesirable stresses and cracks during the drawing process, 
Tuan \textit {et al}. developed two tellurite glasses with high compatibility in thermal and mechanical properties. The two 
tellurite glasses were made of 
${\rm TeO_2}$, ${\rm Li_2O}$, ${\rm WO_3}$ , ${\rm MoO_3}$, ${\rm Nb_2O_5}$ (TLWMN) 
and ${\rm TeO_2}$, ${\rm ZnO}$, ${\rm Na_2O}$, ${\rm La_2O_3}$ (TZNL)
with the refractive index contrast of $\Delta n=0.095$. 
TLWMN and TZNL glass rods were drawn down at 440\textdegree{}C to fibers whose diameters were 100\,\textmu m using a home-designed fiber
drawing tower. In total, 5000 fiber segments of each TLWMN and TZNL glasses were obtained. The 15\,cm long segments were randomly stacked 
with a 50-50 ratio in a TZNL cladding tube whose inner and outer diameters were 10 and 12\,mm, respectively. The fiber was finally drawn to
an all-solid TALOF with the outer diameter of 125\,\textmu m, where the random pixel size was approximately 1.0\,\textmu m.  
The all-solid tellurite-glass TALOF was successfully used to transport the optical image of three vertical slits, with the slit-width as 
small as 14\,\textmu m, over the wavelength range of 1.44-1.60\,\textmu m. The images transported over a 10\,cm length
with high contrast and high brightness.

Finally, we would like to highlight some of on-going as well as past work on phase-separated glasses. The work led by Thomas P Seward III 
of Corning Incorporated in the 1970s on phase-separated glasses resulted in random elongated needle-like structures
after drawing~\cite{seward1974elongation,seward1977some}. The fiber-like glass rods were successfully used 
for image transport and most likely operated based on the TAL principles. More recently, Ballato's group at Clemson University 
has been studying optical fibers that exhibit phase separation during the draw process; i.e., the compounds comprising the core exhibit 
liquid-liquid immiscibility and undergo phase separation in-situ during fiber formation. More specifically, the molten core 
method~\cite{Ballato1} has been used whereby the initial core phase is selected such that it is molten at the temperature where the 
cladding glass softens and draws into fiber. As an example, the rare-earth oxide-silica (${\rm RE}_2{\rm O}_3$-${\rm SiO}_2$) system
is well known to exhibit liquid-liquid immiscibility at high ${\rm SiO}_2$ contents. 
During the fiber draw, silica from the cladding is dissolved by the molten core and becomes incorporated into the core material, forming a silicate melt. 
Thereby the core composition shifts to higher silica concentrations, oftentimes within the liquid-liquid immiscibility dome, leading directly to a 
phase separated core that is quenched into a solid when the fiber cools.
\section{Future directions and conclusions}
Disordered TALOFs demonstrate many novel physical properties, mainly because they allow for localized beam 
transport at any location across the entire cross section of the optical fiber. In general, it is desirable for these
fibers to be designed for the smallest average localized beam diameter. It has been noted that the statistical distribution 
of the localized beam diameters in TALOFs follows a nearly Poisson-like 
distribution~\cite{Mafi-Salman-OPEX-2012,Mafi-Behnam-Scaling-PRB-2016,Mafi-Abaie-OL-2018,mafi-JLT-ArXiv-2019}. 
Therefore, any attempt in reducing the average mode field diameter also reduces the mode-to-mode diameter variations
and results in a more uniform behavior across the fiber cross section. A smaller mode field diameter requires 
a stronger transverse scattering, which is achieved by a higher index contrast, $\Delta n$, as well as a judicious choice
of the random pixel size. The successful experiments using TALOFs have so far been based on $\Delta n \gtrsim 0.1$, 
consistent with the analysis reported in Ref.~\cite{Mafi-Salman-OPEX-2012}. Of course, strongest transverse scattering
is achieved for a 50-50 ratio in the design proposed by De~Raedt \textit {et al}.~\cite{transverse-DeRaedt}. There has been
some unsettled questions and uncertainties regarding a judicious choice of the random pixel 
size~\cite{Mafi-Schirmacher-PRL-2018,Mafi-Abaie-OL-2018,mafi-JLT-ArXiv-2019}--those designs that
target the pixel size to be around half the free-space wavelength, at least for $\Delta n \approx 0.1-0.5$, seem to be 
successful. Of course, in structures that somewhat deviate from that of De~Raedt \textit {et al}. such as the air-glass 
optical fiber by Chen and Li~\cite{chen2014observing}, the rules of design are likely different and must be studied
in detail using appropriate statistical methods~\cite{Mafi-Behnam-Scaling-PRB-2016,Mafi-Abaie-OL-2018}.   

As efforts are made to improve the understanding and design of the transverse microstructure of TALOFs, it is important 
to improve the longitudinal invariance in such fibers to reduce the attenuation. The attenuation for the initial polymer
TALOFs by Karbasi \textit {et al}.~\cite{Mafi-Salman-OL-2012} were in the range of 0.1-1.0\,dB/cm. This relatively large 
attenuation was caused primarily by the exposure of the preliminary fibers in the stack-and-draw procedure to room dust 
during the random mixing process over 3 weeks. Also, the longest piece in which TAL was observed was a 60\,cm segment,
because the fiber thickness varied substantially over the length-scale of one meter or less. Improved attenuation was reported
in glass fibers; e.g., Ref.~\cite{chen2014observing} reported 0.4\,dB/m at 1550\,nm wavelength, and Ref.~\cite{ZHAO:17}
reported 1\,dB/m at visible wavelengths. TALOFS have shown great potential in many applications including in beam
multiplexing~\cite{Mafi-Salman-Multiple-Beam-2013}, image transport~\cite{Mafi-Salman-Nature-2014,zhao2018image,tong2018characterization}, wave-front shaping and sharp
focusing~\cite{Mafi-Marco-Nature-light-focusing-2014}, nonlocal nonlinear behavior~\cite{Mafi-Marco-PRL-Migrating-NL-2014,Mafi-Marco-APL-self-focusing-2014}, 
single-photon data packing~\cite{Mafi-Marco-information-2016}, and random lasers~\cite{Mafi-Behnam-Random-Laser-2017}. Future progress will improve TALOF specifications
for these applications and will open new venues for TALOFs to be utilized.
\section*{Funding}
National Science Foundation (NSF) (1807857) and (1808232).


\begin{thebibliography}{99}
\newcommand{\enquote}[1]{``#1''}
\bibitem{Mafi-Salman-OL-2012}
S.~Karbasi, C.~R. Mirr, P.~G. Yarandi, R.~J. Frazier, K.~W. Koch, and A.~Mafi,
  \enquote{Observation of transverse {A}nderson localization in an optical
  fiber,} {{Opt. Lett.}} \textbf{37}, 2304--2306 (2012).

\bibitem{Anderson1}
P.~W. Anderson, \enquote{Absence of diffusion in certain random lattices,}
  {{Phys. Rev.}} \textbf{109}, 1492--1505 (1958).

\bibitem{Lagendijk-Physics-Today-2009}
A.~Lagendijk, B.~van Tiggelen, and D.~S. Wiersma, \enquote{Fifty-years of
  {A}nderson localization,} {{Physics Today}} \textbf{62},
  24--29 (2009).

\bibitem{sheng2006introduction}
P.~Sheng, \emph{Introduction to wave scattering, localization and mesoscopic
  phenomena} (Springer-Verlag, Berlin, Germany, 2006), 2nd ed.

\bibitem{Soukoulis-1999}
C.~M. Soukoulis and E.~N. Economou, \enquote{Electronic localization in
  disordered systems,} {{Waves in Random Media}}
  \textbf{9}, 255--269 (1999).

\bibitem{matter-waves-2008}
J.~Billy, V.~Josse, Z.~Zuo, A.~Bernard, B.~Hambrecht, P.~Lugan, D.~Cl\'{e}ment,
  L.~Sanchez-Palencia, P.~Bouyer, and A.~Aspect, \enquote{Direct observation of
  {A}nderson localization of matter waves in a controlled disorder,}
  {{Nature}} \textbf{453}, 891--894 (2008).

\bibitem{Lahini-Quantum-Correlation-2010}
Y.~Lahini, Y.~Bromberg, D.~N. Christodoulides, and Y.~Silberberg,
  \enquote{Quantum correlations in two-particle {A}nderson localization,}
  {{Phys. Rev. Lett.}} \textbf{105}, 163905 (2010).

\bibitem{Abouraddy-entangled-2012}
A.~F. Abouraddy, G.~Di~Giuseppe, D.~N. Christodoulides, and B.~E.~A. Saleh,
  \enquote{{A}nderson localization and colocalization of spatially entangled
  photons,} {{Phys. Rev. A}} \textbf{86}, 040302 (2012).

\bibitem{ultrasound-1990}
R.~Weaver, \enquote{{A}nderson localization of ultrasound,}
  {{Wave Motion}} \textbf{12}, 129--142 (1990).

\bibitem{acoustic-PRL-1990}
I.~S. Graham, L.~Pich\'e, and M.~Grant, \enquote{Experimental evidence for
  localization of acoustic waves in three dimensions,}
  {{Phys. Rev. Lett.}} \textbf{64}, 3135--3138 (1990).

\bibitem{elastics-Nat-Phys-2009}
H.~Hu, A.~Strybulevych, J.~H. Page, S.~E. Skipetrov, and B.~A. van Tiggelen,
  \enquote{Localization of ultrasound in a three-dimensional elastic network,}
  {{Nat. Phys.}} \textbf{4}, 945--948 (2008).

\bibitem{John-EM-abs-mobility-edge-1984}
S.~John, \enquote{Electromagnetic absorption in a disordered medium near a
  photon mobility edge,} {{Phys. Rev. Lett.}} \textbf{53},
  2169--2172 (1984).

\bibitem{Chabanov-microwave-2000}
{A.~A.~Chabanov, M.~Stoytchev, A.~Z.~Genack}, \enquote{Statistical signatures
  of photon localization,} {{Nature}} \textbf{404},
  850--853 (2000).

\bibitem{El-Dardiry-microwave-2012}
R.~G.~S. El-Dardiry, S.~Faez, and A.~Lagendijk, \enquote{Snapshots of
  {A}nderson localization beyond the ensemble average,}
  {{Phys. Rev. B}} \textbf{86}, 125132 (2012).

\bibitem{Anderson2}
P.~W. Anderson, \enquote{The question of classical localization a theory of
  white paint?} {{Philos. Mag. B}} \textbf{52}, 505--509
  (1985).

\bibitem{John-Physics-Today-1991}
{S.~John}, \enquote{Localization of light,} {{Phys.
  Today}} \textbf{44}, 32--40 (1991).

\bibitem{SegevNaturePhotonicsReview}
Y.~S. Mordechai~Segev and D.~N. Christodoulides, \enquote{Anderson localization
  of light,} {{Nat. Photonics}} \textbf{7}, 197--204
  (2013).

\bibitem{Mafi-AOP-2015}
A.~Mafi, \enquote{Transverse {A}nderson localization of light: a tutorial,}
  {{Adv. Opt. Photon.}} \textbf{7}, 459--515 (2015).

\bibitem{transverse-Abdullaev}
S.~S. Abdullaev and F.~K. Abdullaev, \enquote{On propagation of light in fiber
  bundles with random parameters,} {{Radiofizika}}
  \textbf{23}, 766--767 (1980).

\bibitem{transverse-DeRaedt}
H.~De~Raedt, A.~Lagendijk, and P.~de~Vries, \enquote{Transverse localization of
  light,} {{Phys. Rev. Lett.}} \textbf{62}, 47--50 (1989).

\bibitem{Schwartz2007}
T.~Schwartz, G.~Bartal, S.~Fishman, and M.~Segev, \enquote{Transport and
  {A}nderson localization in disordered two-dimensional photonic lattices,}
  {{Nature}} \textbf{446}, 52--55 (2007).

\bibitem{Lahini-1D-AL-2008}
Y.~Lahini, A.~Avidan, F.~Pozzi, M.~Sorel, R.~Morandotti, D.~N. Christodoulides,
  and Y.~Silberberg, \enquote{{A}nderson localization and nonlinearity in
  one-dimensional disordered photonic lattices,} {{Phys.
  Rev. Lett.}} \textbf{100}, 013906 (2008).

\bibitem{mafi-JLT-ArXiv-2019}
A.~Mafi, J.~Ballato, K.~W. Koch, and A.~Schulzgen, \enquote{Disordered anderson
  localization optical fibers for image transport-a review,}
  {{arXiv preprint arXiv:1902.00433}}  (2019).

\bibitem{Mafi-Salman-Nature-2014}
S.~Karbasi, R.~J. Frazier, K.~W. Koch, T.~Hawkins, J.~Ballato, and A.~Mafi,
  \enquote{Image transport through a disordered optical fibre mediated by
  transverse {A}nderson localization,} {{Nat Commun}}
  \textbf{5} (2014).

\bibitem{Mafi-Behnam-Random-Laser-2017}
B.~Abaie, E.~Mobini, S.~Karbasi, T.~Hawkins, J.~Ballato, and A.~Mafi,
  \enquote{Random lasing in an {A}nderson localizing optical fiber,}
  {{Light Sci. Appl.}} \textbf{6} (2017).

\bibitem{Mafi-Salman-OPEX-2012}
S.~Karbasi, C.~R. Mirr, R.~J. Frazier, P.~G. Yarandi, K.~W. Koch, and A.~Mafi,
  \enquote{Detailed investigation of the impact of the fiber design parameters
  on the transverse {A}nderson localization of light in disordered optical
  fibers,} {{Opt. Express}} \textbf{20}, 18692--18706
  (2012).

\bibitem{Mafi-Salman-JOVE-2013}
S.~Karbasi, R.~J. Frazier, C.~R. Mirr, K.~W. Koch, and A.~Mafi,
  \enquote{Fabrication and characterization of disordered polymer optical
  fibers for transverse {A}nderson localization of light,}
  {{J. Vis. Exp.}} \textbf{77}, e50679 (2013).

\bibitem{Mafi-Salman-Multiple-Beam-2013}
S.~Karbasi, K.~W. Koch, and A.~Mafi, \enquote{Multiple-beam propagation in an
  {A}nderson localized optical fiber,} {{Opt. Express}}
  \textbf{21}, 305--313 (2013).

\bibitem{Mafi-Salman-OMEX-2012}
S.~Karbasi, T.~Hawkins, J.~Ballato, K.~W. Koch, and A.~Mafi,
  \enquote{Transverse {A}nderson localization in a disordered glass optical
  fiber,} {{Opt. Mater. Express}} \textbf{2}, 1496--1503
  (2012).

\bibitem{chen2014observing}
M.~Chen and M.-J. Li, \enquote{Observing transverse {A}nderson localization in
  random air line based fiber,} in \emph{Photonic and Phononic Properties of
  Engineered Nanostructures IV,}  vol. 8994 (International Society for Optics
  and Photonics, 2014), p. 89941S.

\bibitem{ZHAO:17}
J.~Zhao, J.~E. Antonio-Lopez, R.~A. Correa, A.~Mafi, M.~Windeck, and
  A.~Sch\"{u}lzgen, \enquote{Image transport through silica-air random core
  optical fiber,} in \emph{Conference on Lasers and Electro-Optics,}  (Optical
  Society of America, 2017), p. JTu5A.91.

\bibitem{zhao2018image}
J.~Zhao, J.~E.~A. Lopez, Z.~Zhu, D.~Zheng, S.~Pang, R.~A. Correa, and
  A.~Sch{\"u}lzgen, \enquote{Image transport through meter-long randomly
  disordered silica-air optical fiber,} {{Sci. Rep.}}
  \textbf{8}, 3065 (2018).

\bibitem{zhao2018deep}
J.~Zhao, Y.~Sun, Z.~Zhu, J.~E. Antonio-Lopez, R.~A. Correa, S.~Pang, and
  A.~Schülzgen, \enquote{Deep learning imaging through fully-flexible
  glass-air disordered fiber,} {{ACS Photonics}}
  \textbf{5}, 3930--3935 (2018).

\bibitem{tong2018characterization}
H.~T. Tong, S.~Kuroyanagi, K.~Nagasaka, T.~Suzuki, and Y.~Ohishi,
  \enquote{Characterization of an all-solid disordered tellurite glass optical
  fiber and its near-infrared optical image transport,}
  {{Jpn. J. Appl. Phys.}}  (2018).

\bibitem{seward1974elongation}
T.~P. Seward~III, \enquote{Elongation and spheroidization of phase-separated
  particles in glass,} {{J. Non-Cryst. Solids}}
  \textbf{15}, 487--504 (1974).

\bibitem{seward1977some}
T.~Seward~III, \enquote{Some unusual optical properties of elongated phases in
  glasses,} {{The Physics of Non-Crystalline Solids; Trans
  Tech Publications: Aedermannsdorf, Switzerland}} pp. 342--347 (1977).

\bibitem{Ballato1}
J.~Ballato and P.~Dragic, \enquote{Rethinking optical fiber: New demands, old
  glasses,} {{Journal of the American Ceramic Society}}
  \textbf{96}, 2675--2692 (2013).

\bibitem{Mafi-Behnam-Scaling-PRB-2016}
B.~Abaie and A.~Mafi, \enquote{Scaling analysis of transverse {A}nderson
  localization in a disordered optical waveguide,} {{Phys.
  Rev. B}} \textbf{94}, 064201 (2016).

\bibitem{Mafi-Abaie-OL-2018}
{B. Abaie and A. Mafi}, \enquote{Modal area statistics for transverse
  {A}nderson localization in disordered optical fibers,}
  {{Opt. Lett.}} \textbf{43}, 3834--3837 (2018).

\bibitem{Mafi-Schirmacher-PRL-2018}
W.~Schirmacher, B.~Abaie, A.~Mafi, G.~Ruocco, and M.~Leonetti, \enquote{What is
  the right theory for {A}nderson localization of light? an experimental test,}
  {{Phys. Rev. Lett.}} \textbf{120}, 067401 (2018).

\bibitem{Mafi-Marco-Nature-light-focusing-2014}
M.~Leonetti, S.~Karbasi, A.~Mafi, and C.~Conti, \enquote{Light focusing in the
  {A}nderson regime,} {{Nat. Commun.}} \textbf{5} (2014).

\bibitem{Mafi-Marco-PRL-Migrating-NL-2014}
M.~Leonetti, S.~Karbasi, A.~Mafi, and C.~Conti, \enquote{Observation of
  migrating transverse {A}nderson localizations of light in nonlocal media,}
  {{Phys. Rev. Lett.}} \textbf{112}, 193902 (2014).

\bibitem{Mafi-Marco-APL-self-focusing-2014}
{Leonetti, Marco and Karbasi, Salman and Mafi, Arash and Conti, Claudio},
  \enquote{Experimental observation of disorder induced self-focusing in
  optical fibers,} {{Appl. Phys. Lett.}} \textbf{105},
  171102 (2014).

\bibitem{Mafi-Marco-information-2016}
M.~Leonetti, S.~Karbasi, A.~Mafi, E.~DelRe, and C.~Conti, \enquote{Secure
  information transport by transverse localization of light,}
  {{Sci. Rep.}} \textbf{6} (2016).

\end{thebibliography}
\end{document}